\documentclass[fleqn,10pt]{wlscirep}
\usepackage[utf8]{inputenc}
\usepackage[T1]{fontenc}
\title{Spin independence of the strongly enhanced effective mass in ultra-clean SiGe/Si/SiGe two-dimensional electron system}
\author[1]{M.~Yu.\ Melnikov}
\author[1]{A.~A. Shakirov}
\author[1]{A.~A. Shashkin}
\author[2]{S.-H. Huang}
\author[2]{C.~W. Liu}
\author[3,*]{S.~V. Kravchenko}

\affil[1]{Institute of Solid State Physics, Chernogolovka, Moscow District 142432, Russia}
\affil[2]{Department of Electrical Engineering and Graduate Institute of Electronics Engineering, National Taiwan University, Taipei 106, Taiwan}
\affil[3]{Department of Physics, Northeastern University, Boston, Massachusetts 02115, USA}
\affil[*]{s.kravchenko@northeastern.edu}

\begin{abstract}
\textbf{The effective mass at the Fermi level is measured in the strongly interacting two-dimensional (2D) electron system in ultra-clean SiGe/Si/SiGe quantum wells in the low-temperature limit in tilted magnetic fields.  At low electron densities, the effective mass is found to be strongly enhanced and independent of the degree of spin polarization, which indicates that the mass enhancement is not related to the electrons' spins.  The observed effect turns out to be universal for silicon-based 2D electron systems, regardless of random potential, and cannot be explained by existing theories.
}
\end{abstract}

\begin{document}
\flushbottom
\maketitle

\thispagestyle{empty}

Much interest has been recently directed toward the behaviour of low-disorder, strongly interacting two-dimensional (2D) electron systems.  The interaction strength is characterized by the ratio of the Coulomb and Fermi energies $r_{\text s}=g_{\text v}/(\pi n_{\text s})^{1/2}a_{\text B}$ (here $g_{\text v}$ is the valley degeneracy, $n_{\text s}$ is the electron density and $a_{\text B}$ is the Bohr radius in semiconductor).  In the low-density limit ($r_{\text s}\gtrsim35$), the electrons are expected to form a Wigner crystal \cite{tanatar1989ground,attaccalite2002correlation}.  The Fermi liquid theory is applicable at relatively high electron densities ($r_{\text s}\lesssim1$).  Within this theory, the effective electron mass, $m$, and Land\'e $g$-factor are both enhanced due to spin exchange effects, with renormalization of the $g$-factor being dominant \cite{iwamoto1991static,kwon1994quantum,chen1999exchange}.  Between the Fermi liquid and Wigner crystal, intermediate phases may exist (see, \textit{e.g}., Refs.~\cite{attaccalite2002correlation,spivak2006transport,spivak2010transport,zverev2012microscopic,shaginyan2016strongly}).  A number of theories predict a diverging effective mass near the Wigner crystal or a precursor.  Flattening of the single-particle spectrum at the Fermi level preceded by the diverging effective mass at low electron densities is predicted in Refs.~\cite{zverev2012microscopic,shaginyan2016strongly}.  The effective mass divergence also follows from the renormalization-group theory of the 2D metal-insulator transition \cite{punnoose2005metal}.  Other theoretical approaches envisioning the divergence of the effective mass at low electron densities utilize the analogy between a strongly interacting 2D electron system and He$^3$ \cite{spivak2006transport,spivak2010transport}, Gutzwiller's variational method \cite{brinkman1970application,dolgopolov2002on}, or the dynamical mean-field theory \cite{camjayi2008coulomb}.  Experimentally, a sharp increase of the effective mass and the spin susceptibility, which is proportional to the product $gm$, near the critical density for the metal-insulator transition has been observed in a variety of strongly interacting 2D electron systems: silicon metal-oxide-semiconductor field-effect transistors (MOSFETs) \cite{shashkin2002sharp,anissimova2006magnetization,mokashi2012critical,kuntsevich2015strongly}, GaAs/AlGaAs heterostructures \cite{zhu2003spin,tan2005measurements}, AlAs quantum wells \cite{gokmen2010contrast}, MgZnO/ZnO heterostructures \cite{kasahara2012correlation,falson2015electron,falson2022competing} and SiGe/Si/SiGe quantum wells \cite{melnikov2017indication}, the effect being most pronounced in silicon-based structures.  Particularly, the mass renormalization reaches a factor of about 4 in Si MOSFETs and a factor of about 2 in other 2D systems.  The dramatic increase of the effective mass at the Fermi level with the tendency to diverge is likely to occur near the quantum Wigner crystal in Si MOSFETs \cite{brussarski2018transport} or the flat-band state in ultra-clean SiGe/Si/SiGe quantum wells \cite{melnikov2017indication}.

The renormalization of the effective mass is expected to be strongly sensitive to the polarization of spins, according to the following theoretical approaches.  The effective mass increase with the degree of spin polarization induced by a parallel magnetic field is predicted in Refs.~\cite{spivak2006transport,spivak2010transport} due to the increase of the fraction of crystallites in the mixed state between the Fermi liquid and the Wigner crystal.  In contrast, a decrease of the effective mass with spin polarization is expected from Gutzwiller's variational method, according to Refs.~\cite{dolgopolov2002on,spaek1990almost}.  Calculations of the magnetization within the ring-diagram approximation \cite{zhang2006nonlinear} predict an increase of the spin susceptibility with the magnetic field, whereas a decrease of the effective mass with spin polarization follows from numerical calculations going beyond the random-phase approximation \cite{asgari2009effective}.  In measurements of the effective electron mass in a dilute two-valley 2D electron system in silicon MOSFETs in tilted magnetic fields, the enhanced value $m$ was found, within the experimental uncertainty, to be independent of the degree of spin polarization \cite{shashkin2003spin}, which is in disagreement with the above-mentioned theories.  Measurements of the spin susceptibility in Si MOSFETs were reported in which non-monotonic 10\% variations of the spin susceptibility with parallel magnetic field were observed \cite{pudalov2004on,pudalov2021magnetic}; note that the observed weak change of the spin susceptibility can be due to the change of the $g$-factor with spin-independent mass.  On the other hand, the application of the parallel magnetic field at low electron densities in AlAs quantum wells resulted in a decrease of the effective mass even below the band value in the single-valley case and an about 15\% increase of the effective mass in the two-valley case \cite{gokmen2010contrast,padmanabhan2008effective}.  The contradictory experimental results call for experiments to be performed in alternative 2D electron systems.

In this paper, we report measurements of the effective mass at the Fermi level in the two-valley 2D electron system in ultra-clean SiGe/Si/SiGe quantum wells in tilted magnetic fields.  This electron system is distinguished in that (i)~the peak electron mobility is almost two orders of magnitude higher than that in the least-disordered Si MOSFETs (110~m$^2$/Vs and 3~m$^2$/Vs, correspondingly); and (ii) the behaviours of the effective mass at the Fermi level and the energy-averaged effective mass are qualitatively different at low electron densities, indicating the band flattening at the Fermi level \cite{melnikov2017indication,zverev2012microscopic,shaginyan2016strongly}.  The effective mass is extracted from the temperature dependence of the weak-field Shubnikov-de~Haas (SdH) oscillations in the low-temperature limit.  It is enhanced by a factor of $\gtrsim2$ at low electron densities, \textit{i.e.}, in the strongly-interacting regime where the ratio between Coulomb and Fermi energies reaches a factor of about 20.  We find that within the experimental uncertainty, the value of the effective mass does not depend on the degree of spin polarization, which points to a spin-independent origin of the effective mass enhancement.  The observed effect turns out to be universal for silicon-based 2D electron systems, regardless of random potential, and is in contradiction with existing theories.

\begin{figure*}
\scalebox{.405}{\includegraphics[width=\columnwidth]{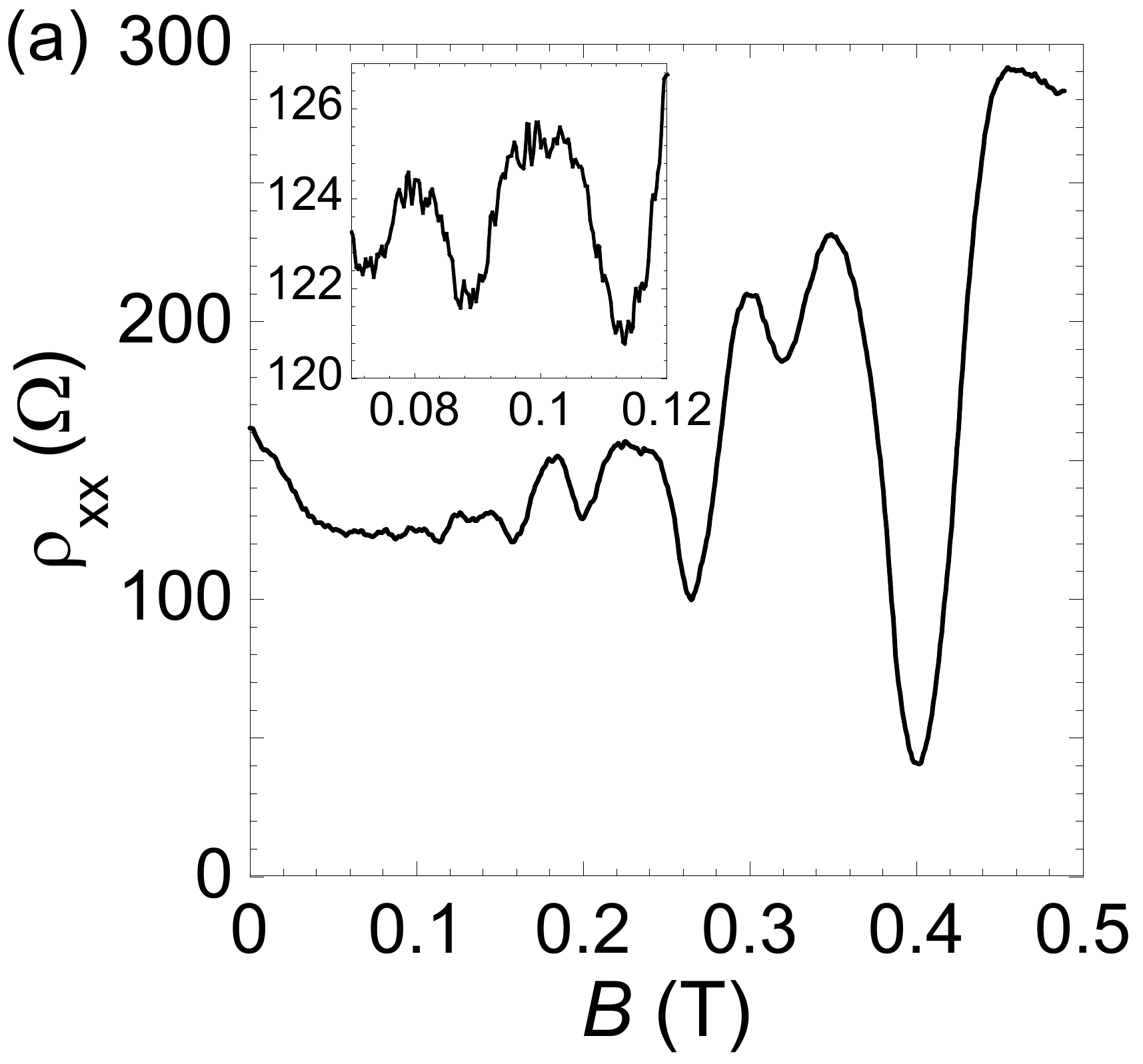}}
\scalebox{.415}{\includegraphics[width=\columnwidth]{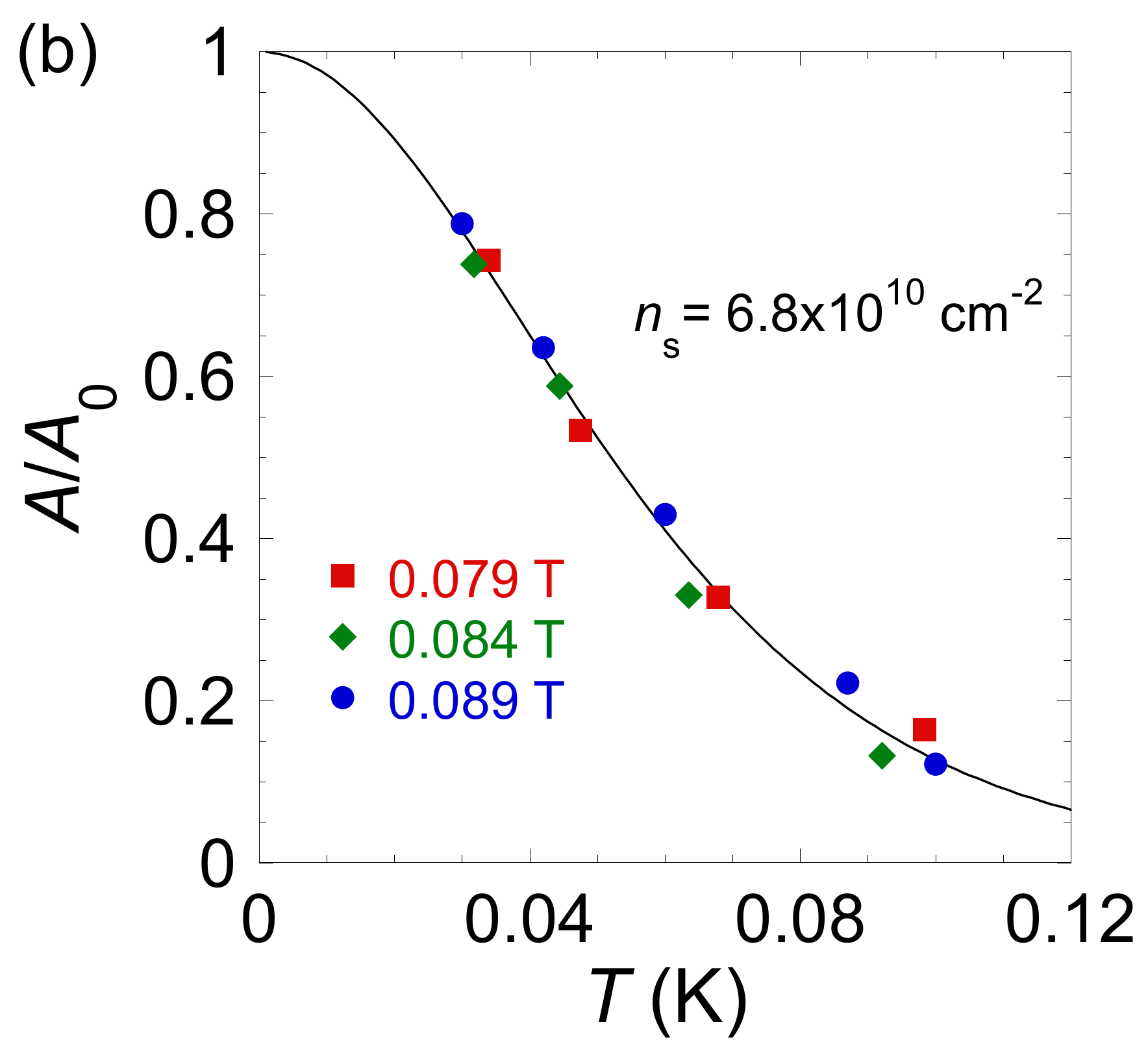}}
\caption{(a) Shubnikov-de~Haas oscillations in perpendicular magnetic fields at $T=30$~mK and $n_{\text s}=3.97\times10^{10}$~cm$^{-2}$.  A fragment of the weak-field SdH oscillations averaged over 32 traces is shown in the inset.  (b)~Change of the amplitude of the weak-field SdH oscillations in perpendicular magnetic fields with the temperature at $n_{\text s}=6.8\times10^{10}$~cm$^{-2}$ and $B_1=$ 0.079~T (squares), $B_2=$ 0.084~T (diamonds) and $B_3=$ 0.089~T (circles).  The values of $T$ for the data obtained in $B_1$ and $B_2$ are multiplied by $B_3/B_1$ and $B_3/B_2$, correspondingly.  The solid line is a fit using Eq.~(\ref{A}).}
\label{fig1}
\end{figure*}

Typical SdH oscillations of the resistivity $\rho_{\text{xx}}(B)$ are shown in Fig.~\ref{fig1}~(a), with the inset displaying a fragment of the weak-field (sinusoidal) oscillations.  The experiments were demanding because at electron densities $\gtrsim10^{10}$~cm$^{-2}$, measuring $\rho_{\text{xx}}(B)$ oscillations required weak magnetic fields $\sim0.1$~T and low temperatures $\lesssim0.1$~K.  To improve the signal-to-noise ratio, the traces $\rho_{\text{xx}}(B)$ were recorded up to 70 times and then averaged.  Temperature dependence of the amplitude, $A$, of the weak-field SdH oscillations for the normalized resistance, $R_{xx}/R_0$ (where $R_0$ is the average resistance), is displayed in Fig.~\ref{fig1}~(b).  To determine the effective mass, we fit the data for $A(T)$ using the Lifshitz-Kosevich formula \cite{lifshits1955on,isihara1986density}
\begin{equation}
A(T)=A_0\frac{2\pi^2k_{\rm B} T/\hslash\omega_{\rm c}}{\sinh(2\pi^2k_{\rm B}T/\hslash \omega_{\rm c})}, \;\; A_0=4\exp(-2\pi^2k_{\rm B} T_{\rm D}/\hslash\omega_{\rm c}),
\label{A}
\end{equation}
where $\omega_c=eB_\perp/mc$ is the cyclotron frequency, $B_\perp$ is the perpendicular component of the magnetic field and $T_{\rm D}$ is the Dingle temperature related to the level width through the expression $T_{\rm D}=\hslash/2\pi k_{\rm B}\tau_{\rm q}$ (here $\tau_{\rm q}$ is the quantum scattering time).  Damping of the SdH oscillations with temperature may be influenced by the temperature-dependent $\tau_{\rm q}$.  However, in our experiments, possible corrections to the effective mass due to the temperature dependence of $\tau_{\rm q}$ are within the experimental uncertainty and do not exceed 10\%; note that the temperature corrections to $m$ and $\tau_{\rm q}$, calculated in Ref.~\cite{adamov2006interaction}, are small and can be ignored in our case.  The amplitude of the SdH oscillations follows the calculated curve down to the lowest achieved temperatures, which confirms that the electrons were in good thermal contact with the bath and were not overheated.

\begin{figure*}
\scalebox{.4175}{\includegraphics[width=\columnwidth]{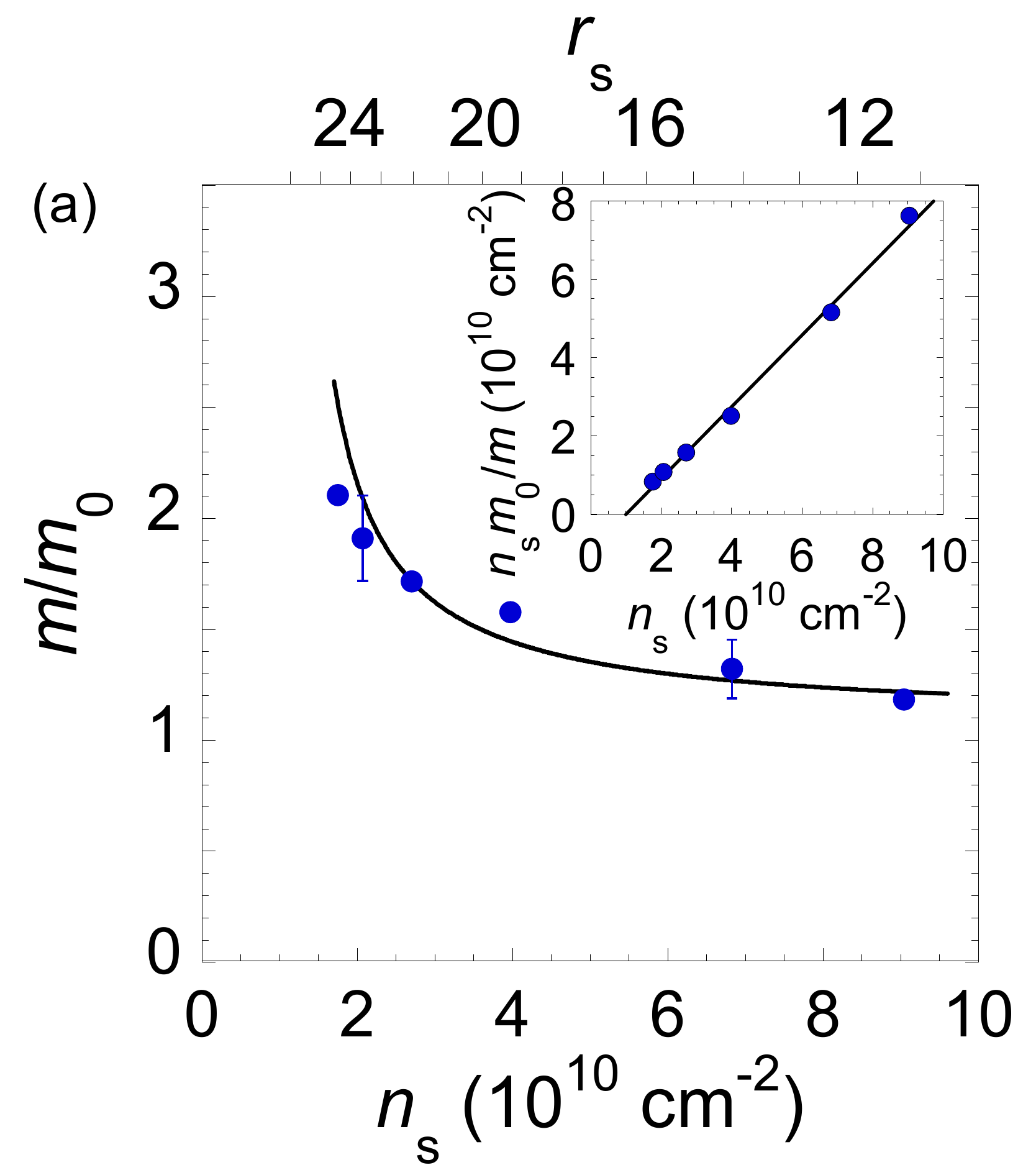}}
\scalebox{.395}{\includegraphics[width=\columnwidth]{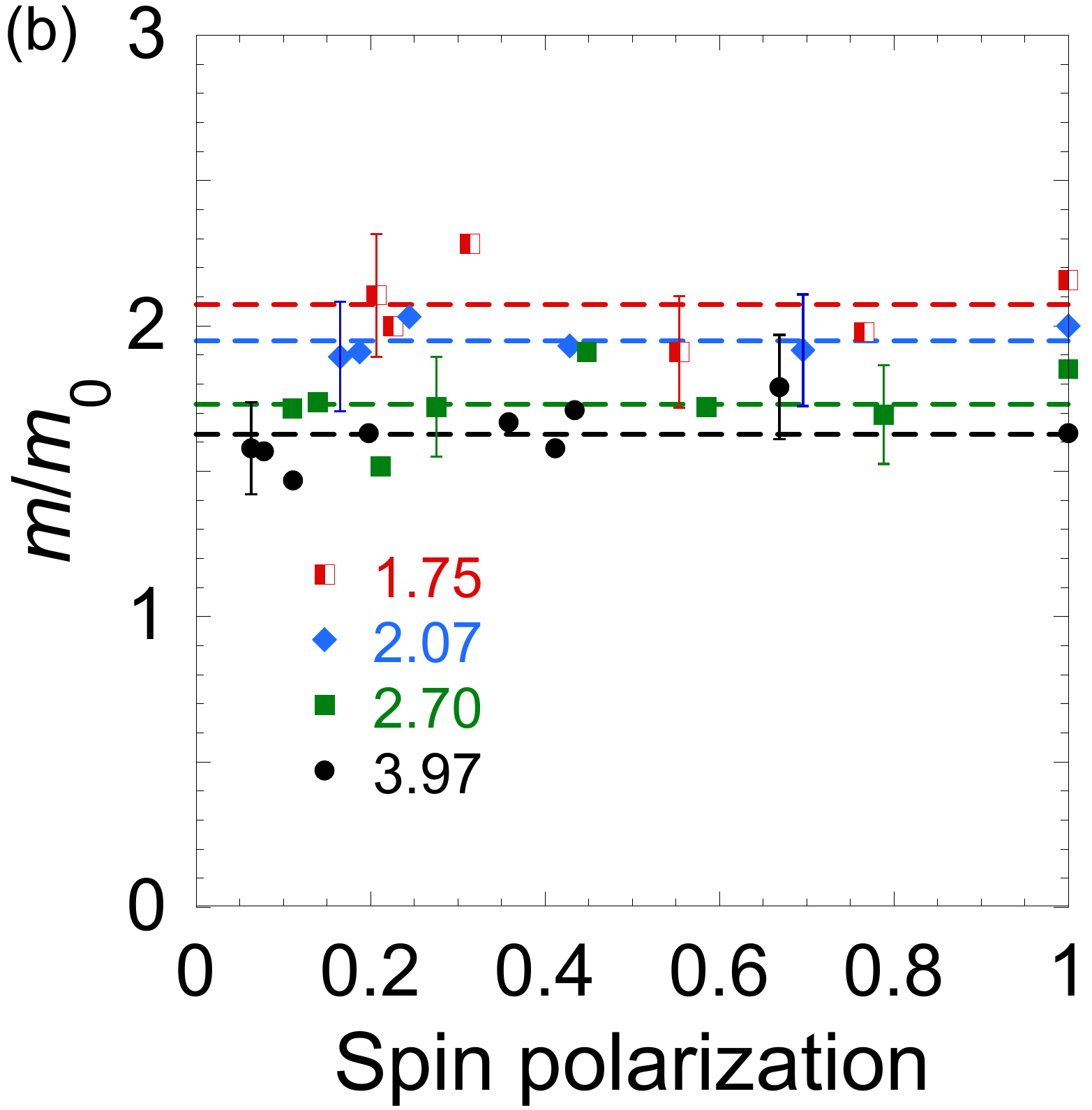}}
\caption{(a) The effective mass as a function of electron density determined by SdH oscillations in perpendicular magnetic fields.  Also shown are the $r_{\text s}$ values on the top axis.  The solid line corresponds to the one in the inset.  The inset shows the dependence $n_{\text s}m_0/m$ on the electron density.  The solid line is a linear fit that extrapolates to zero at $n_{\text s}\approx1.0\times10^{10}$~cm$^{-2}$.  (b)~The effective mass \textit{vs}.\ the degree of spin polarization for the following electron densities in units of $10^{10}$~cm$^{-2}$: 1.75 (half-filled squares), 2.07 (diamonds), 2.70 (squares), and 3.97 (circles).  The dashed horizontal lines are guides to the eye.}
\label{fig2}
\end{figure*}

In Fig.~\ref{fig2}~(a), we show the effective mass, normalized by $m_0$ (where $m_0=0.19\;m_{\text e}$ is the effective mass for noninteracting electrons and $m_{\text e}$ is the free electron mass), as a function of the electron density determined by SdH oscillations in perpendicular magnetic fields; the $r_{\text s}$ values have been calculated using the simple formula mentioned above (\textit{cf.} Refs.~\cite{melnikov2019quantum,shashkin2021metal}).  In measurements, appreciably lower values of $n_{\text s}$ have been reached as compared to previous experiments on this electron system.  The effective mass sharply increases with decreasing $n_{\text s}$, in agreement with previously obtained results \cite{melnikov2017indication}.  As seen from the inset to the figure, the inverse effective mass extrapolates linearly to zero at a density $n_{\text s}\approx1.0\times10^{10}$~cm$^{-2}$.  This is consistent with the results of measurements of the product $gm$ at the Fermi level, in which case $g\approx g_0$ (where $g_0=2$ is the $g$-factor for noninteracting electrons) \cite{melnikov2019quantum}.  It is worth noting that the similar dependence of the inverse effective mass on electron density in Si MOSFETs extrapolates to zero at a density $n_{\text s}\approx8\times10^{10}$~cm$^{-2}$, and both critical densities in SiGe/Si/SiGe quantum wells and Si MOSFETs correspond to approximately the same $r_{\text s}$ value, according to Ref.~\cite{melnikov2019quantum}.

\begin{figure*}
\scalebox{0.44}{\includegraphics[width=\columnwidth]{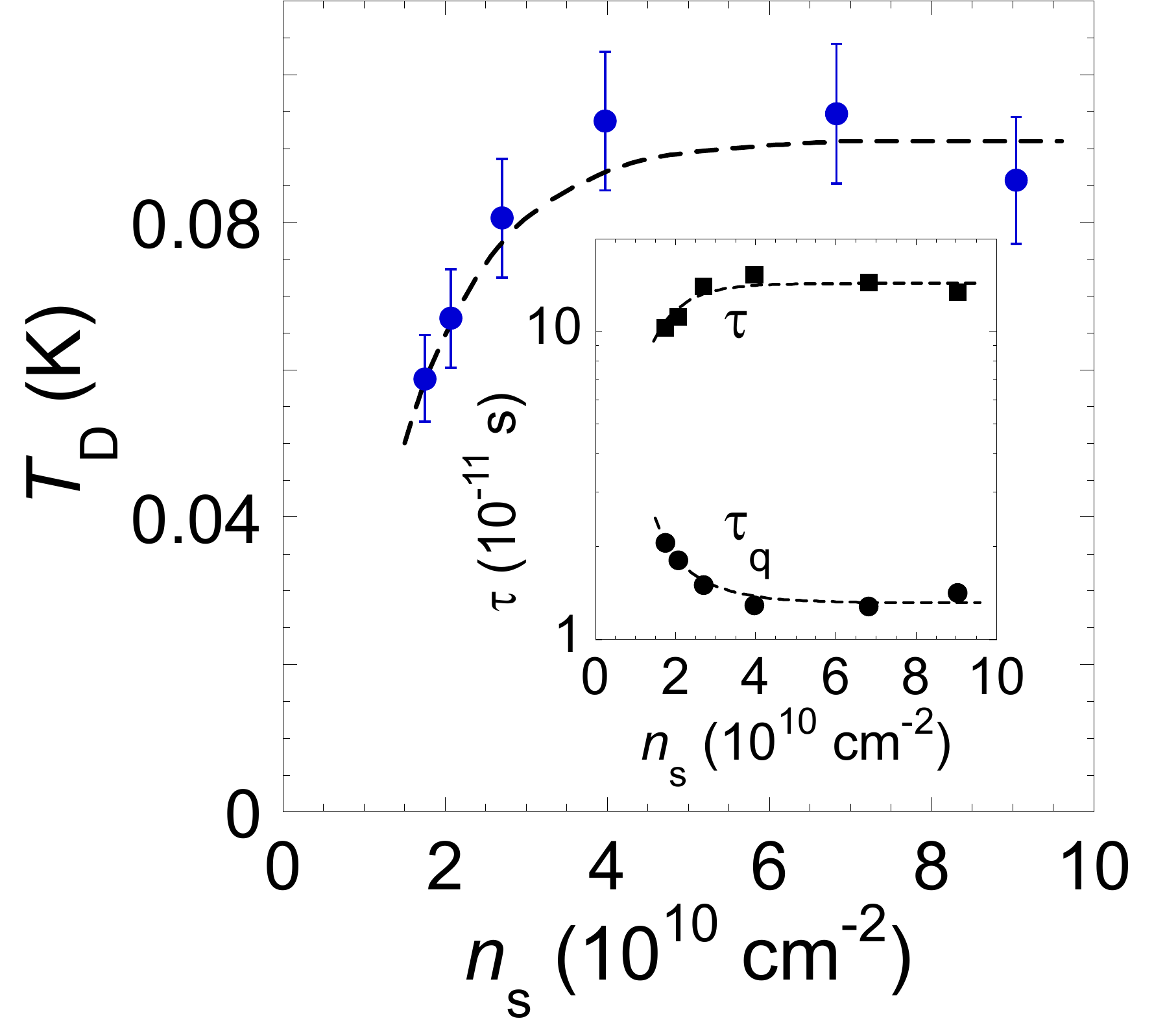}}
\caption{Dingle temperature as a function of the electron density extracted from SdH oscillations in perpendicular magnetic fields. The dashed line is a guide to the eye.  The inset shows the density dependence of the corresponding quantum scattering time $\tau_{\text q}$ (circles) and the transport scattering time $\tau$ (squares) recalculated from the electron mobility. The dashed lines are guides to the eye.}
\label{fig3}
\end{figure*}

A strong enhancement of the effective mass at the Fermi level in the limit of low electron densities is theoretically expected to be related to spin effects, as mentioned above.  To probe a possible influence of spins on the effective mass enhancement, we used tilted magnetic fields to produce a parallel magnetic field component, $B_\parallel$, that couples primarily to the electrons' spins; note that the orbital effects for $B_\parallel$ due to the finite thickness of the 2D electron system can be neglected in our samples in the range of magnetic fields studied \cite{melnikov2017unusual}.

The main result of the paper is shown in Fig.~\ref{fig2}~(b): within the experimental uncertainty, the enhanced effective mass at the Fermi level does not depend on the degree of spin polarization, $p=(B_\parallel^2+B_\perp^2)^{1/2}/B_{\text c}$ (where $B_{\text c}$ is the parallel magnetic field of the complete spin polarization corresponding to the magnetoresistance saturation, determined as described in Ref.~\cite{melnikov2017indication}).  This indicates that the origin of the mass enhancement in our samples has no relation to the spin exchange effects.  This finding is in agreement with the results obtained in Si MOSFETs \cite{shashkin2003spin}.

In Fig.~\ref{fig3}, we plot the Dingle temperature, $T_{\text D}$, as a function of the electron density extracted from SdH oscillations in perpendicular magnetic fields.  Within the experimental uncertainty, the Dingle temperature changes weakly with electron density at $n_{\text s}\gtrsim4\times10^{10}$~cm$^{-2}$, while at lower electron densities, the value $T_{\text D}$ drops with decreasing $n_{\text s}$.  It is worth noting that a similar decrease of the Dingle temperature at low electron densities was observed in Si MOSFETs \cite{shashkin2003spin}; in that electron system, the quantum scattering time $\tau_{\text q}$ and the transport scattering time, $\tau$, determined from the electron mobility are approximately equal to each other, and the short-range random potential is realized. The decrease of $T_{\text D}$ with decreasing electron density is in agreement with the narrowing of the cyclotron resonance line observed at low $n_{\text s}$ in Si MOSFETs \cite{cheng1990anomalies}.  In the inset to Fig.~\ref{fig3}, we compare the density dependences of the corresponding quantum scattering time $\tau_{\text q}$ and the transport scattering time $\tau$.  The transport scattering time exceeds by far $\tau_{\text q}$, indicating the dominant long-range random potential in the samples;  at the same time, the presence of a short-range random potential and electron backscattering is inferred from the pronounced metallic temperature dependence of the resistance in zero magnetic field \cite{melnikov2015ultra}.  In addition, both scattering times converge somewhat at low electron densities.  One can indeed expect a decrease of the ratio $\tau/\tau_{\text q}$ at low $n_{\text s}$ because for the same transferred wavevector, the decrease of the Fermi wavevector with decreasing electron density should lead to a larger-angle electron scattering and smaller ratios $\tau/\tau_{\text q}$ \cite{dolgopolov2003remote}.  Both scattering times are expected to vanish in the insulating phase (for more on this, see Ref.~\cite{shashkin2002sharp}).

The independence of the strongly enhanced effective mass at the Fermi level of the degree of spin polarization, observed in ultra-clean SiGe/Si/SiGe quantum wells, is concurrent with the results obtained in Si MOSFETs, although the random potentials in these electron systems are very different.  We arrive at a conclusion that the observed effect is universal for silicon-based 2D electron systems, regardless of random potential.

Although the theoretically predicted divergence of the effective mass at low densities due to the electron interaction effects \cite{brinkman1970application,dolgopolov2002on,punnoose2005metal,spivak2006transport,camjayi2008coulomb,spivak2010transport,zverev2012microscopic,shaginyan2016strongly} agrees with the experiments, the observed spin polarization independence of the effective mass cannot be explained by existing theories.  Particularly, within the models of Refs.~\cite{spivak2006transport,zhang2006nonlinear,spivak2010transport}, the effective mass is expected to increase with spin polarization, while the prediction of Refs.~\cite{dolgopolov2002on,spaek1990almost,asgari2009effective} is the opposite.  Note that in contrast to the irrelevance of spin effects in the observed behaviour of the effective mass, the exchange and correlation effects manifest themselves in the observed increase in the critical density for the metal-insulator transition in spin-polarizing magnetic fields (see, \textit{e.g.}, Refs.~\cite{dolgopolov2017spin,melnikov2020metallic}).  Since two or more competing mechanisms or states are in play, the problem of the interaction-enhanced effective mass in spin-polarizing magnetic fields is very complicated and requires considerable theoretical efforts.

In summary, we have found that at low electron densities in the strongly interacting 2D electron system in ultra-clean SiGe/Si/SiGe quantum wells, the effective mass at the Fermi level is strongly enhanced and independent of the degree of spin polarization, which indicates that the mechanism underlying the effective mass enhancement is unrelated to the electrons' spins.  The observed effect turns out to be universal for silicon-based 2D electron systems, regardless of random potential, and is not explained by existing theories.

\section*{Methods}

The samples were double-gate ultra-high mobility SiGe/Si/SiGe quantum wells similar to those described in detail in Refs.~\cite{melnikov2015ultra,melnikov2017unusual,dolgopolov2021valley}.  The approximately 15~nm wide silicon (001) quantum well is sandwiched between Si$_{0.8}$Ge$_{0.2}$ potential barriers.  The peak electron mobility in these samples reached 110~m$^2$/Vs.  The samples were patterned in Hall-bar shapes with the distance between the potential probes of 150~$\mu$m and width of 50~$\mu$m using standard photo-lithography.  Measurements were carried out in an Oxford TLM-400 dilution refrigerator.  Data were taken by a standard four-terminal lock-in technique in a frequency range 1--11~Hz at currents 0.03--4~nA in the linear regime of response.

\subsection*{Data availability.} The data that support the findings of this study are available from the corresponding author upon reasonable request.


We gratefully acknowledge discussions with V. Dobrosavljevi\'c, V.~A. Khodel, M. Shayegan and B. Tanatar.  The ISSP group was supported by RSF Grant No.\ 22-22-00333.  The NTU group acknowledges support by the Ministry of Science and Technology, Taiwan (Project No.\ 112-2218-E-002-024-MBK).  S.V.K. was supported by NSF Grant No.\ 1904024.\vspace{2mm}\\
The authors declare no competing financial or non-financial interests.\vspace{2mm}\\
Correspondence and requests for materials should be addressed to S.V.K.\\ (email: s.kravchenko@northeastern.edu).

\section*{Author contributions}
This project was conceived, planned and executed by M.~Yu. Melnikov, A.~A. Shashkin and S.~V. Kravchenko. Data were taken by M.~Yu. Melnikov, A.~A. Shakirov and A.~A. Shashkin. The SiGe/Si/SiGe wafers were grown by S.-H. Huang and C.~W. Liu and processed by M.~Yu. Melnikov  and A.~A. Shashkin.  The data analysis was made by M.~Yu. Melnikov, A.~A. Shashkin and S.~V. Kravchenko.  The manuscript was composed by A.~A. Shashkin and S.~V. Kravchenko.  All the authors reviewed the manuscript.\\
\newpage

\end{document}